\documentclass[twocolumn, prl,superscriptaddress,showpacs,noshowkeys,reprint,floatfix]{revtex4-1}
\pdfpageattr {/Group << /S /Transparency /I true /CS /DeviceRGB>>}

\usepackage{graphicx}
\usepackage{amsmath}
\usepackage[version=3]{mhchem}
\usepackage[utf8]{inputenc}
\usepackage[T1]{fontenc}
\usepackage[finnish,english]{babel}
\usepackage{verbatim}


\begin{document}

\title{Intermolecular contrast in atomic force microscopy images without intermolecular bonds}
\author{Sampsa K. H\"{a}m\"{a}l\"{a}inen }
\affiliation{Department of Applied Physics, Aalto University School of Science, PO Box 15100, 00076 Aalto, Finland}
\author{Nadine van der Heijden}
\affiliation{Condensed Matter and Interfaces, Debye Institute for Nanomaterials Science,
Utrecht University, PO Box 80000, 3508 TA Utrecht, the Netherlands}
\author{Joost van der Lit}
\affiliation{Condensed Matter and Interfaces, Debye Institute for Nanomaterials Science,
Utrecht University, PO Box 80000, 3508 TA Utrecht, the Netherlands}
\author{Stephan den Hartog}
\affiliation{Condensed Matter and Interfaces, Debye Institute for Nanomaterials Science,
Utrecht University, PO Box 80000, 3508 TA Utrecht, the Netherlands}
\author{Peter Liljeroth}
\email[]{peter.liljeroth@aalto.fi}
\affiliation{Department of Applied Physics, Aalto University School of Science, PO Box 15100, 00076 Aalto, Finland}
\author{Ingmar Swart}
\email[]{i.swart@uu.nl}
\affiliation{Condensed Matter and Interfaces, Debye Institute for Nanomaterials Science,
Utrecht University, PO Box 80000, 3508 TA Utrecht, the Netherlands}

\date{\today}

\begin{abstract}
Intermolecular features in atomic force microscopy (AFM) images of organic molecules have been ascribed to intermolecular bonds. A recent theoretical study [P. Hapala et al., Phys. Rev. B 90, 085421 (2014)] showed that these features can also be explained by the flexibility of molecule-terminated tips. We probe this effect by carrying out AFM experiments on a model system that contains regions where intermolecular bonds should and should not exist between close-by molecules. Intermolecular features are observed in both regions, demonstrating that intermolecular contrast cannot be directly interpreted as intermolecular bonds.
\end{abstract}

\maketitle

The use of molecule-modified tips in non-contact atomic force microscopy (AFM) has enabled the visualization of the chemical structure of molecules with unprecedented resolution \cite{Gross28082009,Gross:2010aa,deOteyza2013}.  Molecule-terminated tips have also been used to probe the bond orders in conjugated molecules and to map the charge distribution inside a molecule \cite{Gross14092012,Mohn:2012aa}. Recently, even intermolecular features assigned to hydrogen bonds have been reported \cite{Zhang01112013,Sweetman:2014aa}. According to the IUPAC definition \cite{arunan2011definition}, hydrogen bonds primarily have an electrostatic origin and may include some covalent character and other attractive interactions. It is not clear why these should yield significant repulsive contrast in AFM \cite{hapala2014mechanism,Sweetman:2014aa}. {As AFM images with submolecular resolution were obtained with molecule-modified tips, the tip flexibility has to be considered for a quantitative understanding of the results \cite{PhysRevLett.106.046104,Gross14092012,doi:10.1021/nn500317r,Weymouth07032014,PhysRevB.89.205407,Moll2014}.

Ab initio calculations [e.g. by density functional theory (DFT)] of the CO tip-substrate system are hampered by the fact that the exact atomic structure of the metal tip behind the CO molecule is unknown. This affects, for example, the calculated lateral force constant. On the other hand, it has been demonstrated that a molecular mechanics approach is sufficient for a quantitative understanding of the contrast formation in the AFM images \cite{doi:10.1021/nn500317r}. The AFM contrast over the bonds is caused by the presence of a saddle surface setup by the spherical potentials from the nearby atoms. As the tip-sample distance is decreased, the CO flexibility causes an apparent sharpening of the bonds \cite{Gross14092012,doi:10.1021/nn500317r}. Similar mechanism should operate irrespective of the origin of the saddle surface in the interaction potential landscape. For example, intramolecular features that do not correspond to chemical bonds have been observed \cite{Pavlicek2012}. Recent computational work by the Jelinek group suggests that the intermolecular features observed in AFM are not related to actual hydrogen bonds, but are rather caused by the CO flexibility and the shape of the potential landscape between the molecules \cite{hapala2014mechanism}. The same mechanism was shown to be responsible for the observed contrast in scanning tunnelling hydrogen microscopy (STHM) and in inelastic tunneling probe microscopy \cite{Temirov2008,Weiss2010,Chiang23052014,hapala2014mechanism,arxiv.1409.3405}.

Despite the compelling arguments in favor of the intermolecular features having been caused by the CO flexibility, the systems considered thus far also had hydrogen bonds in the positions of the enhanced AFM contrast. This makes it difficult to establish the origin of intermolecular contrast in AFM images. In this letter, we focus on a molecular system, where four bis(para-pyridyl)acetylene (BPPA) molecules form a tetramer stabilized by hydrogen bonds. This results in two nitrogen atoms from neighboring molecules being forced close together without chemical or hydrogen bonds being formed between them. We experimentally show that an apparent intermolecular bond shows up in AFM images where no bond exists. We corroborate these experiments using a molecular mechanics model and quantitatively match the observed contrast with the expected response caused by the flexibility of the CO molecule at the tip apex.

The BPPA molecules were synthesized according to Ref. \cite{Champness1999}. Samples were prepared by evaporating the BPPA molecules from a Knudsen cell-type evaporator onto a Au(111) single crystal, cleaned by sputtering/annealing cycles. After preparation, the sample was inserted into a low-temperature STM/AFM ($T = 4.8$ K, Omicron LT-STM/qPlus AFM), housed within the same ultrahigh vacuum system (base pressure $\sim10^{-10}$ mbar). We used a qPlus sensor with a resonance frequency $f_0$ of 24454 Hz, a quality factor of > 10000, spring constant $k=1800$ N/m, and a peak-to-peak oscillation amplitude of $\sim 1.7$ Å (amplitude $A=0.85$ Å). Picking up an individual carbon monoxide molecule to the tip apex was carried out as described previously \cite{PhysRevLett.80.2004,Gross28082009}. For the constant-height AFM images, the tip-sample distance was typically decreased by a few tens of pm (indicated in the figure captions) w.r.t. the STM set-point ($V_\mathrm{bias} = 0.1$ V, $I = 10$ pA) after switching off the feedback. To eliminate creep and minimize drift, the tip was allowed to stabilize for 12 hours before AFM images were acquired (recorded with $V_\mathrm{bias}=0$ V).

Upon adsorption on the Au(111) substrate, the BPPA molecules self-assemble into several different structures. Fig. \ref{fig1}a shows one of the more common structures. It consists of BPPA tetramers (Fig. \ref{fig1}b), which are held together by \ce{C\bond{-}H\cdots N} hydrogen-bonds between the electronegative pyridinic nitrogens and the hydrogen atoms of the pyridine rings. A ball-and-stick model of these tetramers is given in Fig. \ref{fig1}c. Within the tetramer, the hydrogen bonds (red dashed lines in Fig. \ref{fig1}c) force the nitrogens of the two opposing BPPA molecules close to each other ($\approx$ 3 Å).

We can exclude the presence of coordinating Au atoms in the middle of the tetramers as the distance between the molecules is much smaller than expected for two times a typical Au-N bond \cite{mohr}. In addition, if present, a Au atom should be visible in the AFM image, even if it is located below the plane of the molecules \cite{Albrecht2013}.

\begin{figure}
\includegraphics[width=\linewidth]{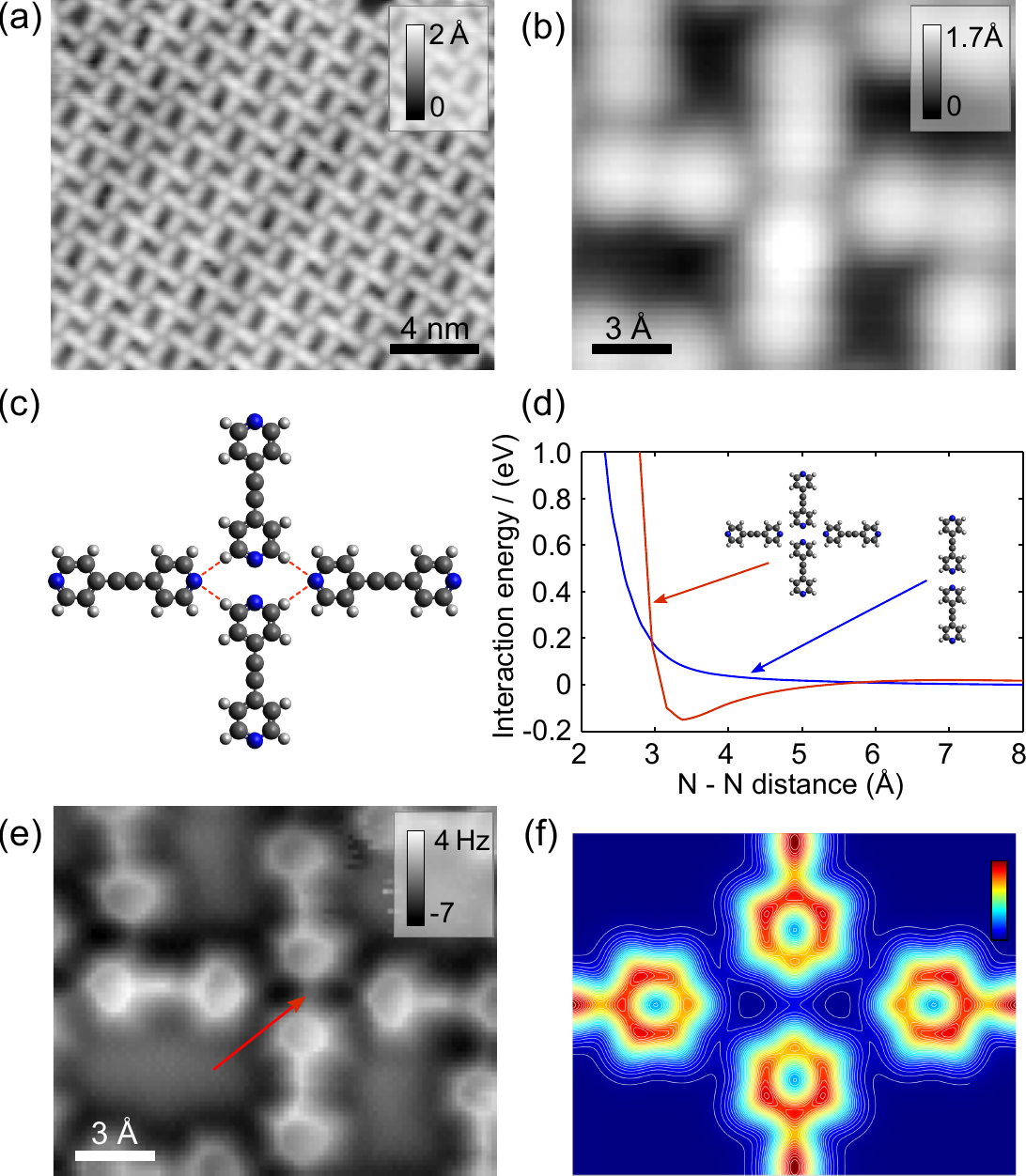}
\caption{(Color online) (a) Overview STM image of the self-assembled BPPA molecules. (b) STM image of a single BPPA tetramer. (c) Schematic of the tetramer. (d) Energy-distance curve for two BPPA molecules (blue line) and for a tetramer (red line). (e) AFM image of the tetramer taken with a CO terminated tip showing apparent intermolecular bonds. (f) Total electron density 3.1 Å above the molecular plane given by DFT.}
\label{fig1}
\end{figure}

We have analyzed the structure of the tetramer by DFT calculations of the molecules in the absence of the substrate. Omission of the substrate is justified as the molecules are expected to interact weakly with the Au(111) surface \cite{PhysRevLett.102.176102,Lit2013}. We used dispersion corrected GGA (PBE-D3) and hybrid functionals (B3LYP-D3) in combination with a TZ2P basis set as implemented in ADF \cite{ADF}. The energy vs. distance curve of a dimer (Fig. \ref{fig1}d) shows pure repulsive behavior, consistent with the expected repulsive interaction between the electron lone pairs of the pyridinic nitrogens. In contrast, for the tetramer we find a binding energy of -151 meV (PBE-D3). Hence, the tetramer represents a stable configuration. The distance between the nitrogen atoms is in reasonable agreement with the experimental values (3.3 Å vs. 3.0 Å).

The constant-height AFM image of a BPPA tetramer taken with a CO terminated tip (Fig. \ref{fig1}e) shows weak contrast on the \ce{C-H\cdots N} bond, similar to the recently published results on imaging hydrogen bonds with AFM \cite{Zhang01112013,Sweetman:2014aa}. However, there is also contrast in the region between the opposing nitrogen atoms (indicated by the arrow in Fig. \ref{fig1}e), despite the absence of a bond between these atoms.

It has been proposed that the atomic scale contrast can be modelled by considering only the total electron density at the position of the AFM tip apex \cite{1367-2630-12-12-125020}. However, as discussed previously \cite{Sweetman:2014aa}, this is not sufficient to explain the AFM response even in the case of a rigid tip apex. Additional effects, such as the depletion of the charge density due to the tip-sample interaction may need to be taken into account \cite{Sweetman:2014aa}. The discrepancy between the total electron density and the AFM response is also seen in our data. Fig. \ref{fig1}f shows the total electron density 3.1~\AA~above the plane of the molecules. As expected, the electron density in the region between the molecules is much lower than in between atoms. In the conventional model of the AFM imaging with CO terminated tips \cite{1367-2630-12-12-125020}, this should translate in a much weaker intermolecular contrast, contrary to what we observe experimentally. We will show below that the contrast in the actual measurement can be understood by considering the bending of the CO molecule at the tip apex \cite{doi:10.1021/nn500317r,hapala2014mechanism}.

We use a model based on molecular mechanics \cite{doi:10.1021/nn500317r}, similar to the model used by Hapala et al. \cite{hapala2014mechanism}. The tip is modeled in two parts; a macroscopic sphere representing the metallic bulk tip and a CO molecule which is allowed to move on a lever attached to the apex of the sphere. The macroscopic sphere and the substrate under the molecules were treated on a continuum level, while the CO and the molecules were treated atomistically as shown schematically in Fig. \ref{fig2}a. We use a Lennard-Jones 9-6 potential for all the interatomic interactions with the parameters rescaled from the 12-6 potential from Ref. \cite{1998_Mackerell} such that the position and depth of the potential minimum remained constant. We tested different forms of the repulsive interaction; the exact form has only a very minor effect on the simulated AFM response. While electrostatic interactions have been shown to be relevant with CO terminated tips \cite{Schneiderbauer2014,APL2014}, we neglect them in the present case as the vertical dipole of the CO is not expected to interact strongly with the horizontal dipole of the hydrogen bond. In addition, electrostatic interactions should be heavily screened by the metallic substrate and tip. Finally, electrostatic force components would only cause the location of the saddle point in the interaction potential surface to shift along the hydrogen bond and would not have an effect on the AFM contrast across the bond. The CO was relaxed self-consistently at each point before calculating the forces. The frequency shift $\Delta f$ at a given height was calculated taking into account the finite oscillation amplitude used in the experiment \cite{giessibl2001direct}.

\begin{figure}
\includegraphics[width=\linewidth]{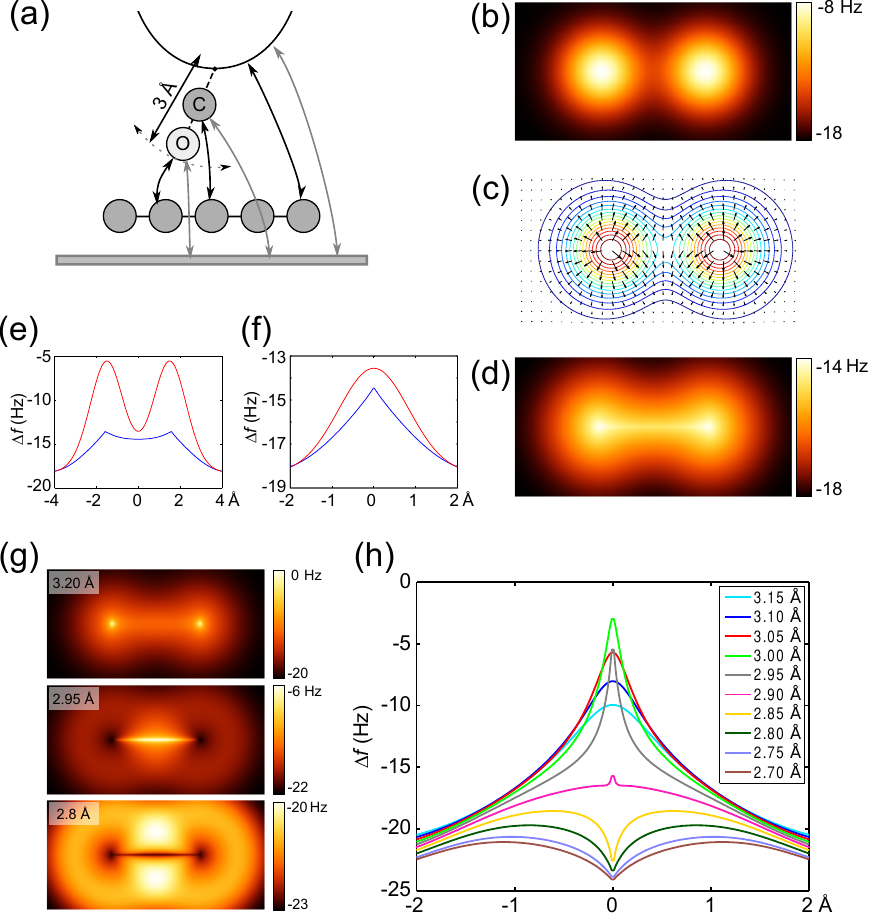}
\caption{(Color online) (a) Illustration of the tip model used in the simulation. The arrows refer to the different force components that either are functions of the $(x,y,z)$-coordinates (black) or only depend on the $z-$coordinate (gray). (b) Simulated constant-height AFM image (tip height 2.9 Å) with a rigid CO tip of two carbon atoms 3 Å apart. (c) CO potential on top of the atoms (contour) and the bending of the CO in the simulation. (d) Simulated constant-height AFM image with a flexible CO tip ($k_\mathrm{tip}=0.6$ N/m) at the same tip height as in panel (b). (e) and (f) cross-sections along and across the apparent bond, respectively. (g) $\Delta f$ images calculated at the low amplitude limit at different heights. (h) $\Delta f$ cross-sections across the apparent bond between the atoms at the low-amplitude limit. All tip-sample distances are measured from the oxygen of the CO at the lowest point of oscillation.}
\label{fig2}
\end{figure}

Fig. \ref{fig2}b shows a calculated constant-height AFM image of two carbon atoms 3 Å apart with a rigid CO molecule on the tip apex. As expected, the resulting $\Delta f$ image shows two spherically symmetric maxima centered on top of the atoms. The interaction potential felt by the CO molecule is plotted in Fig. 2c. Letting the CO respond to the lateral forces by bending (quiver plot in Fig. 2c) has a dramatic influence on the simulated AFM image (Fig. \ref{fig2}d). The two atoms produce a saddle surface between them, which causes the CO to bend away from the line connecting the atoms. The bending of the CO causes a decrease in the measured $\Delta f$ signal. Thus, the sharp contrast along the line connecting the atoms is the result of the reduced repulsion away from the symmetry line rather than an increased repulsion between the atoms (due to, e.g., increased electron density). This is illustrated in Fig. \ref{fig2}e and f, which show cross sections of the $\Delta f$ surfaces across and along a line between the atoms for a rigid (red) and a flexible (blue) CO tip.

This is the essential effect behind the very sharp contrast obtained in AFM with CO terminated tips on molecules where the bonds between the atoms appear un-physically sharp \cite{Gross14092012,doi:10.1021/nn500317r}. The sharpening will happen on any saddle-like potential surface: it does not require actual electron density between the atoms, i.e. a bond. This is clearly shown in the simulation in Fig. \ref{fig2}d where no bond exists as the CO interaction with the atoms is modeled by the L-J potential. Consequently, the CO bending will pick up the shape of the potential energy surface between two molecules, and symmetry lines will show with enhanced contrast. This means that intermolecular contrast cannot be taken as a proof of the existence of an intermolecular bond \cite{hapala2014mechanism}.

\begin{figure*}[ht!]
\includegraphics[width=\linewidth]{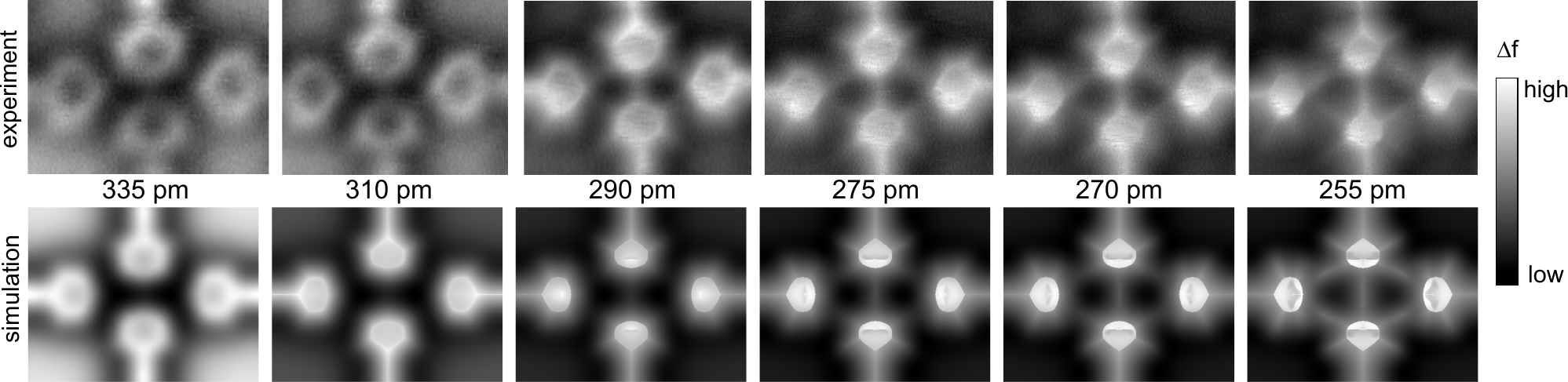}
\caption{(top row) Experimental constant-height AFM images with a CO tip taken at different heights on top of the tetramer junction showing the appearance of both  \ce{C-H\cdots N} and \ce{N-N} intermolecular contrast at close tip-sample distances. (bottom row) Simulated constant-height AFM images with a flexible CO tip ($k_\mathrm{tip}=0.6$ N/m) at the given heights showing the appearance of the same intermolecular contrast. The relative height scale is same in the experimental and simulated images with the simulated height of 385 pm matching the tunneling conditions of 0.1 V / 10 pA. The heights correspond to the lowest point of the tip oscillation.}
\label{fig3}
\end{figure*}

In order to verify this effect experimentally, we have studied the evolution of the intermolecular features within the BPPA tetramer as a function of the tip-sample distance. Figure \ref{fig3} shows a set of constant-height AFM images taken with a CO terminated tip at different heights above the BPPA tetramer junction with a comparison to the images produced by our simple CO tip model. In neither the experiments nor in the simulations are intermolecular bond-like features observed for large tip-sample distances, as the CO never reaches the repulsive regime. Taking the tip closer, lines start to appear both between the two opposing nitrogens and on the actual hydrogen bonds holding the tetramer together. The contrast first appears between the two opposing nitrogens as they are closer to one another and hence the saddle surface is formed there first. Upon approaching the tip further, contrast also appears in the region of the hydrogen bonds that hold the tetramer together. At the smallest tip-sample distance, the CO bending starts to dictate the contrast formation. Surprisingly, the contrast on the intermolecular lines become almost indistinguishable from the contrast on the acetylene moieties connecting the pyridine rings in the BPPA molecule, both in experiment and simulation.

The sharp contrast produced by the bending of the CO molecule can easily be incorrectly interpreted as overly high resolution. In reality, the bending of the CO sets a limit to the resolution that can be obtained in AFM with flexible tips. As can be seen in Figs. \ref{fig2}g and h, the simulated low-amplitude $\Delta f$ signal starts to level out and invert when the tip is pushed further in. The height at which this happens is defined by the stiffness of the tip. This results in a loss of contrast between repulsion maxima (e.g. on top of atoms) and other areas of repulsion when integrating over the tip oscillation amplitude. This eventually renders all saddle surfaces and atoms equally bright in the $\Delta f$ image irrespective of the magnitude of the repulsion, or electron density. At this point the contrast formation in AFM is dictated by the lateral stiffness of the CO and is no longer related to the magnitude of the tip-sample interaction.

This work and the work by Hapala et al. \cite{hapala2014mechanism} raises an important question:  To what extent is tip bending responsible not just for the intermolecular, but also for the intramolecular contrast in planar molecules? In other words: can AFM image bonds, or is the technique only sensitive to the potential energy landscape originating from (spherically symmetric) potentials of the atoms in the molecule? A covalent bond is a region of enhanced electron density between two atoms. If electron density would be the only contribution to the contrast, it is very difficult to understand the similar contrast between atoms (where the electron density is much higher) and bonds. Indeed, tip relaxations are essential to reproduce experimentally observed intramolecular contrast \cite{hapala2014mechanism,Pavlicek2012}. This suggests that tip flexibility also plays a dominant role in imaging intramolecular bonds.

As discussed in the introduction, electrostatic interactions and small changes in electron density are not expected to result in repulsive contrast in AFM images. In order to see hydrogen bonds experimentally, one would need to be able to resolve the extra electron density caused by the bonding. This would require a quantitative estimate of the electron density on the atoms in order to extrapolate onto the region of the hydrogen bonds. This is experimentally very demanding as the tip flexibility will cause loss of contrast on the atoms at these small tip-sample distances.

In conclusion, we present an AFM measurement of BPPA tetramers using a CO terminated tip. Identical intermolecular contrast appears both on top of the hydrogen bonds and between two pyridinic nitrogens which do not bond. We show that the CO bending causes two effects which enhance the apparent intermolecular features in AFM. The CO bends away from the ridges in the saddle surface of the interaction potential, which produces sharp lines between nearby atoms. At the same time, bending away from the actual atoms decreases the $\Delta f$ signal on top of the molecules, which increases the relative intensity on the intermolecular features. This means that the contrast on both real and apparent bonds is mostly a result of the bending of the probe molecule on the AFM tip. Hence, intermolecular contrast in AFM images does not necessarily represent intermolecular bonds.

\begin{acknowledgments}
We thank Adam Foster, Peter Spijker, and Leo Gross for discussions. This research was supported by the European Research Council (ERC-2011-StG No. 278698 "PRECISE-NANO"), the Academy of Finland (Centre of Excellence in Low Temperature Quantum Phenomena and Devices No. 250280), and the Netherlands Organization for Scientific Research (NWO-ECHO-STIP grant No. 717.013.003).
\end{acknowledgments}


%

\end{document}